\documentclass[11pt,twoside]{article}
\usepackage{asp2010}

\resetcounters

\begin{document}

\title{WorldWide Telescope in Research and Education}
\author{Alyssa Goodman$^1$, Jonathan Fay$^2$, August Muench$^1$, Alberto Pepe$^1$, Patricia Udompraseret$^1$, and Curtis Wong$^2$
\affil{$^1$Harvard-Smithsonian Center for Astrophysics, Cambridge, MA}
\affil{$^2$Microsoft Research, Redmond, WA}}

\begin{abstract}
The WorldWide Telescope computer program, released to researchers and the public as a free resource in 2008 by Microsoft Research, has changed the way the ever-growing Universe of  online astronomical data is viewed and understood.    The WWT program can be thought of as a scriptable, interactive, richly visual browser of  the multi-wavelength Sky as we see it from Earth, and of the  Universe as we would travel within it.  In its web API format, WWT is being used as a service to display professional research data.  In its desktop format, WWT works in concert (thanks to SAMP and other  IVOA standards) with more traditional research applications such as ds9, Aladin and TOPCAT.  The WWT Ambassadors Program (founded in 2009) recruits and trains astrophysically-literate volunteers (including retirees) who use WWT as a teaching tool in online, classroom, and informal educational settings. Early quantitative studies of WWTA indicate that student experiences with WWT enhance science learning dramatically.  Thanks to the wealth of data it can access, and the growing number of services to which it connects, WWT is now a key linking technology in the \textit{Seamless Astronomy} environment we seek to offer researchers, teachers, and students alike.  
\end{abstract}

\section{Introduction}
The ``WorldWide Telescope" was originally envisioned, by Jim Gray and Alex Szalay in 2001, as ``an Archetype for Online Science" \citep{SzalayGray2001}. ÊWhen the program called ``WordWide Telescope" was released by Microsoft Research in 2008, it was received by the press primarily as an amazing new tool for outreach--offering access to the world's best astronomical imagery and expertise to all. ÊSince its release, though, the free WWT software has become both an essential new piece of the research ecosystem and the amazing educational tool the press perceived \citep{GoodmanWong2009}. ÊAll versions of WWT discussed in this article are available  at no cost for non-commercial use at \url{worldwidetelesope.org}.

\section{WWT in Astronomical Research}

\begin{figure} []
\centering
\includegraphics{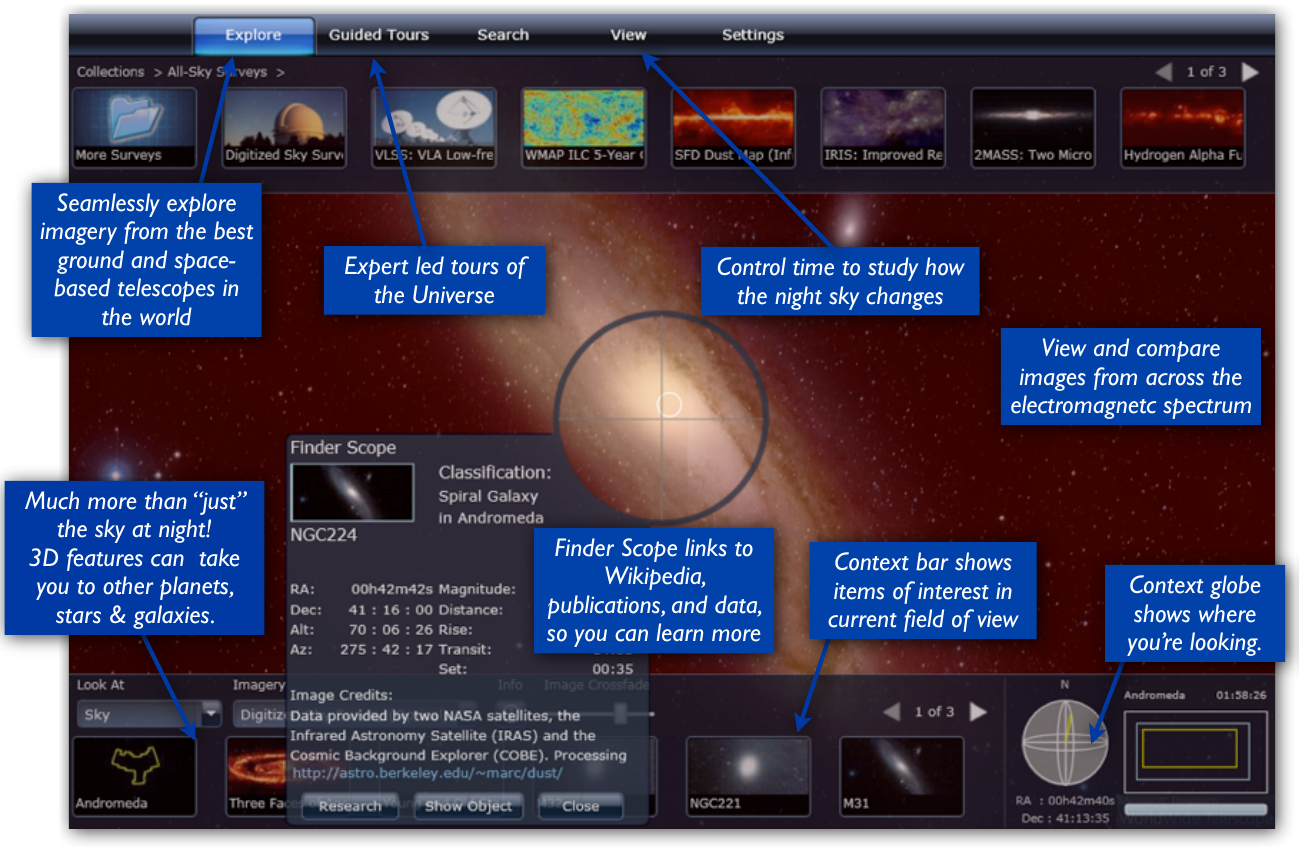}
\caption{Annotated screen shot of the WorldWide Telescope application, as it appears in either its Windows desktop version or as a Silverlight-based Web application.  Notice in particular the ÒFinder ScopeÓ functionality associated with the (movable) cross-hair.}
\label{fig1}
\end{figure}

The full version of WWT (WWT-desktop) runs as a standalone desktop application in Windows (either on a Windows-only system, or on a Mac running a Windows Virtual Machine).   In addition, WWT runs within a web browser on any machine capable of running Silverlight (e.g. almost any Mac or PC in use today).     On the web, users and developers have a choice of: 1) a menu-driven version of WWT that looks identical to WWT-desktop for Sky-based work; or 2) an API offering a fully-functional data-viewing window plus added functionality (e.g. ``Finder Scope") as-desired (see Figure \ref{fig1}).

At present, WWT-desktop is SAMP-compliant, and near-future compliance with WebSAMP (Taylor 2012, this volume) is planned.  In concert with SAMP, and other VO tools, WWT is a fantastic viewer for manipulating, overlaying, and cross-matching image and catalog-based information.  ÊIn its API form, WWT makes a powerful visualization tool, and a clear example of how it can be used to show survey coverage and user-selectable data layers is given at the COMPLETE Survey's data-coverage page, at \url{www.worldwidetelescope.org/COMPLETE/WWTCoverageTool.htm}.

The fast, smooth, panning and zooming in and out possible within all forms of WWT, in combination with its ``context globe" (see Figure \ref{fig1}), scale indicators, and built-in all-sky multi-wavelength views offers a contextual perspective on astronomical data that has not been possible before.  This  kind of context is typically missing in professionals' views of their data, and its presence allows for better understanding of the potential interactions amongst various physical processes.

The context WWT offers is now also accessible from within ADS Labs (\url{adslabs.org}), under the {\it{``SIMBAD Objects"}} facet there, making the linkages between WWT and ADS now truly two-way.  (WWT has, since its initial release, offered a direct link to ADS articles about any point in the sky via the {\it{``Research, Information, Look up publications on ADS"}} option in the Finder Scope, and direct links to CDS/SIMBAD are accessible similarly via {\it{``Research, Information, Look up on SIMBAD."}})  Soon, the just-funded NASA ADS All Sky Survey will employ WWT as one of a few all-sky viewers capable of showing image holdings extracted from ADS articles in-context on the sky, filterable by subject, object, author, time, and more (Pepe et al. 2012, this volume).

\section{WWT in Teaching and Learning}

In education, WWT is being used at all age and expertise levels. ÊThe WorldWide Telescope Ambassadors Program (WWTA), founded as a Harvard-Microsoft collaboration in 2009, trains astronomy experts to use WWT in both informal and formal (classroom) environments \citep{Udomprasertetal2011}. Ambassadors are trained to create ``Tours" within WWT, and to facilitate the use of the program and the Tours.  Tours are interactive paths through WWT's Sky or 3D content designed to make a point.  Sample tours include {\it Galileo's New Order}, which explains how Galileo's discovery of Jupiter's moons led to the adoption of our current heliospheric view of our Solar System, and {\it John Huchra's Universe}, a tribute to John Huchra that explains the significance of redshift surveys to our understanding of our Universe.  Tours created and/or vetted by experts are served at the WWT Ambassadors website (\url{wwtambassadors.org}, and many are also accessible from the ``Tours" menu tab within the program).  The WWTA website provides a faceted education-friendly view of all Tour content, as well as an area where students and teachers working on their own Tours can share them.  

\begin{figure} [h]
\centering
\includegraphics{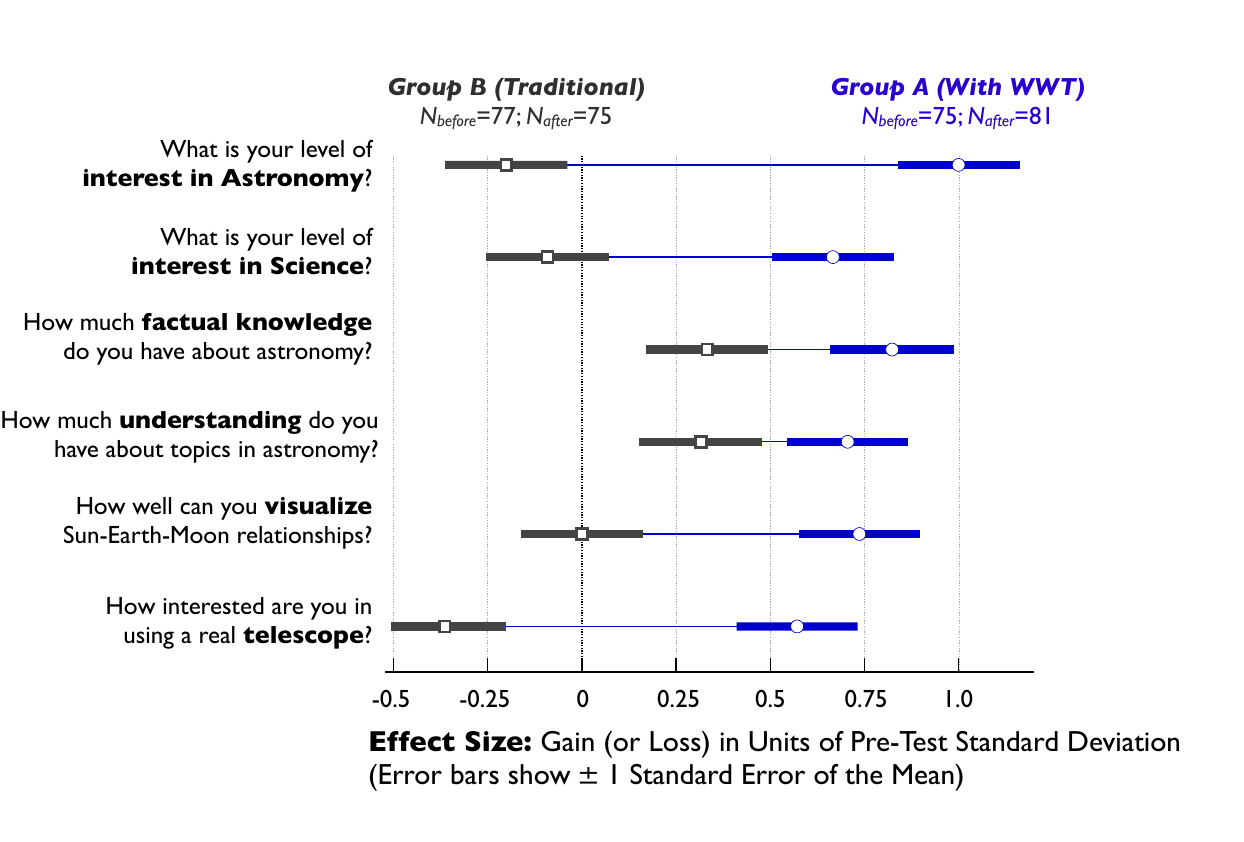}
\caption{Results from the Clarke Middle School Pilot of the WorldWide Telescope Ambassadors Program (2010).}
\label{fig2}
\end{figure}

In the 2010 pilot of WWTA at Clarke Middle School in Lexington, MA, two groups of $\sim75$ sixth-grade students  were studied: the ``treatment" group participated in WWTA and the other group was a control, with access only to the standard curriculum.  Treatment students all created their own Tours, in groups of three or four.  Ambassadors in the classroom facilitated the Tour creation, mostly by pointing students to additional online astrophysical resources, as these 11-year-olds typically did not need much help learning to use the program.  Each group of students was surveyed before and after six-weeks of astronomy study, and the comparative results are shown in Figure \ref{fig2}.  WWTA had dramatic effects across the board, increasing knowledge, understanding, and interest in science, and even increasing student interest in using real, physical, telescopes! \citep{Goodmanetal2011}.   In 2011, the WWTA  Program was successfully expanded to several more Boston-area schools, and we (in collaboration with Stephen Strom of NOAO) are actively seeking funding to facilitate much-asked-for US and international expansion of the Program.

In higher-education, WWT is now part of several University courses and labs, and we (in collaboration with  Edwin Ladd of Bucknell University) have additional proposals pending to develop curricular materials appropriate to these higher levels. Since the best of today's university students are the near-future's most important researchers, the ``educational" environment of universities actually offers the greatest potential for expanding WWT's use in {\it research} in the near-term future.

Astronomy has been aptly called a ``gateway drug" for STEM (Science, Technology, Engineering and Math) learning, and as such we feel a responsibility to Êexpand the WWTA site and programs to include all ages of learners, from pre-K to retirees.

\section{WWT in the Future}
WWT is a key part of a larger program based at the Harvard-Smithsonian Center for Astrophysics called ``Seamless Astronomy" (for a list of collaborators, see \url{projects.iq.harvard.edu/seamlessastronomy/}).  The vision of Seamless Astronomy is shared by many ADASS participants:  astronomical research tools should interoperate so well that boundaries between data archives, countries, and program(s) functionality all but disappear.   WWT has demonstrated how several forms of the same tool, accessing many different data archives, and connecting ``seamlessly" (thanks to SAMP and other IVOA standards) to many other tools and services, can make astronomy researchÐand STEM educationÐeasier, and so much more fun.

\bibliographystyle{asp2010}
\bibliography{O15}
\end{document}